\documentclass[aps,prd,twocolumn,superscriptaddress]{revtex4}
\usepackage{epsfig,epsf}
\usepackage{amsmath}
\usepackage{amsthm}
\usepackage{amsfonts}
\usepackage{amssymb}
\usepackage{dsfont}
\usepackage{multirow}
\usepackage{appendix}
\usepackage{slashed}
\usepackage[active]{srcltx}
\usepackage{psfrag}

\begin{document}

\title{Heavy four-quark mesons $bc\overline{b}\overline{c}$: Scalar particle}
\date{\today}
\author{S.~S.~Agaev}
\affiliation{Institute for Physical Problems, Baku State University, Az--1148 Baku,
Azerbaijan}
\author{K.~Azizi}
\affiliation{Department of Physics, University of Tehran, North Karegar Avenue, Tehran
14395-547, Iran}
\affiliation{Department of Physics, Do\v{g}u\c{s} University, Dudullu-\"{U}mraniye, 34775
Istanbul, T\"{u}rkiye}
\affiliation{Department of Physics and Technical Sciences, Western Caspian University,
Baku, AZ 1001, Azerbaijan}
\author{H.~Sundu}
\affiliation{Department of Physics Engineering, Istanbul Medeniyet University, 34700
Istanbul, T\"{u}rkiye}

\begin{abstract}
Parameters of the heavy four-quark scalar meson $T_{\mathrm{bc\overline{b}%
\overline{c}}}$ with content $bc \overline{b}\overline{c}$ are calculated by
means of the sum rule method. This structure is considered as a
diquark-antidiquark state built of scalar diquark and antidiquark
components. The mass and current coupling of $T_{\mathrm{bc\overline{b}%
\overline{c}}}$ are evaluated in the context of the two-point sum rule
approach. The full width of this tetraquark is estimated by taking into
account two types of its possible strong decay channels. First class
includes dissociation of $T_{\mathrm{bc\overline{b}\overline{c}}} $ to
mesons $\eta_c\eta_{b}$, $B_{c}^{+}B_{c}^{-}$, $B_{c}^{\ast +}B_{c}^{\ast -}$
and $B_{c}^{+}(1^3P_{0})B_{c}^{\ast-}$. Another type of processes are
generated by annihilations $\overline{b}b \to \overline{q}q$ of constituent $%
b$-quarks which produces the final-state charmed meson pairs $D^{+}D^{-}$, $%
D^{0} \overline{D}^{0}$, $D^{*+}D^{*-}$, and $D^{*0}\overline{D}^{*0}$.
Partial width all of these decays are found using the three-point sum rule
method which is required to calculate strong couplings at corresponding
meson-meson-tetraquark vertices. Predictions obtained for the mass $m=(12697
\pm 90)~\mathrm{MeV}$ and width $\Gamma[T_{\mathrm{bc\overline{b}\overline{c}%
}}]=(142.4 \pm 16.9)~ \mathrm{MeV}$ of this state are compared with
alternative results, and are useful for further experimental investigations
of fully heavy resonances.
\end{abstract}

\maketitle


\section{Introduction}

\label{sec:Intro}
Due to recent experimental achievements of different collaborations \cite%
{LHCb:2020bwg,Bouhova-Thacker:2022vnt,CMS:2023owd}, fully heavy four-quark
mesons composed of only $c$ and $b$ quarks became objects of intensive
studies. The four $X$ resonances observed in di-$J/\psi $ and $J/\psi \psi
^{\prime }$ mass distributions by the LHCb, ATLAS, and CMS Collaborations
are first candidates to such exotic mesons. These structures with the masses
in the interval $6.2-7.3~\mathrm{GeV}$ are presumably scalar particles $cc%
\overline{c}\overline{c}$ composed of four valence $c$ quarks.

Features of $X$ resonances were investigated in the context of different
models and theoretical approaches \cite%
{Zhang:2020xtb,Albuquerque:2020hio,Yang:2020wkh,Becchi:2020mjz,
Becchi:2020uvq,Wang:2022xja,Faustov:2022mvs,Niu:2022vqp,Dong:2022sef,
Yu:2022lak,Kuang:2023vac,Wang:2023kir,Dong:2020nwy,Liang:2021fzr}. The fully
charmed scalar four-quark mesons in the diquark-antidiquark and molecule
pictures were considered in our publications as well \cite%
{Agaev:2023wua,Agaev:2023ruu,Agaev:2023gaq,Agaev:2023rpj}. By applying the
QCD two- and three-point sum rule methods, we computed masses and widths of
such states. Having compared obtained predictions with experimental data, we
concluded that the lightest structure $X(6200)$ is supposedly a molecule $%
\eta _{c}\eta _{c}$ \cite{Agaev:2023ruu}, whereas $X(6600)$ maybe is the
diquark-antidiquark state built of axial-vector ingredients \cite%
{Agaev:2023wua}. The superposition of a tetraquark with a pseudoscalar
diquark and antidiquark components and hadronic molecule $\chi _{c0}\chi
_{c0}$ has parameters which agree with parameters of the state $X(6900)$
\cite{Agaev:2023ruu,Agaev:2023gaq}. The heaviest resonance $X(7300)$ can be
considered as an admixture of the hadronic molecule $\chi _{c1}\chi _{c1}$
and first radial excitation of $X(6600)$ \cite{Agaev:2023rpj}.

An interesting class of fully heavy tetraquarks which deserve detailed
analysis are particles $bb\overline{c}\overline{c}$/$cc\overline{b}\overline{%
b}$. These exotic mesons bear two units of electric charge and under certain
conditions can be strong-interaction stable states. Tetraquarks $bb\overline{%
c}\overline{c}$/$cc\overline{b}\overline{b}$ with different quantum numbers
were explored in numerous publications \cite%
{Wu:2016vtq,Li:2019uch,Wang:2019rdo,Liu:2019zuc}, in which were made
contradictory conclusions about their stability. The four-quark mesons $bb%
\overline{c}\overline{c}$ with spin-parities $J^{\mathrm{P}}=0^{\pm }$,\ $%
1^{\pm }$, and $2^{+}$ were explored in our works \cite%
{Agaev:2023tzi,Agaev:2024pej,Agaev:2024pil,Agaev:2024xdc}. We calculated the
masses of the diquark-antidiquark states with these quantum numbers and
found that none of them are strong-interaction stable particles. We also
evaluated widths of these states by considering their kinematically allowed
decay channels.

All-heavy exotic mesons $bc\overline{b}\overline{c}$ form another group of
interesting particles. These states were explored intensively in the
literature as well \cite%
{Faustov:2022mvs,Wu:2016vtq,Liu:2019zuc,Chen:2019vrj,Bedolla:2019zwg,Cordillo:2020sgc,Weng:2020jao,Yang:2021zrc,Hoffer:2024alv}%
, where masses of structures $bc\overline{b}\overline{c}$ with different
spin-parities were calculated by applying alternative approaches. Thus, in
Ref.\ \cite{Faustov:2022mvs} analyses were done in the context of the
relativistic quark model, whereas the authors in Refs.\ \cite%
{Yang:2021zrc,Hoffer:2024alv} used the QCD sum rule and relativistic
four-body Faddeev-Yakubovsky approaches, respectively. Predictions obtained
in these works for parameters of the tetraquarks $bc\overline{b}\overline{c}$
differ from each another considerably. Thus, the mass of the scalar particle
$J^{\mathrm{PC}}=0^{++}$ was found equal to $12.813~\mathrm{GeV}$, $%
12.28-12.46~\mathrm{GeV}$, and $10.72~\mathrm{GeV}$, respectively. The large
differences between these results are evident which makes actual new
detailed studies of the tetraquarks $bc\overline{b}\overline{c}$.

In present article, we consider the scalar diquark-antidiquark state $T_{%
\mathrm{bc\overline{b}\overline{c}}}$ with the quark content $bc\overline{b}%
\overline{c}$ built of scalar diquark components. We are going to perform
comprehensive analysis of this structure and calculate its mass and full
width. To determine the mass and current coupling of $T_{\mathrm{bc\overline{%
b}\overline{c}}},$ we make use of the two-point sum rule (SR) approach \cite%
{Shifman:1978bx,Shifman:1978by}. It turns out that this exotic meson lies
above the $\eta _{b}\eta _{c}$, $B_{c}^{+}B_{c}^{-}$, $B_{c}^{\ast
+}B_{c}^{\ast -}$ and $B_{c}^{+}(1^{3}P_{0})B_{c}^{\ast -}$ thresholds and
hence dissociates to these two-meson final states. Apart from that, due to $b%
\overline{b}$ annihilation to light quarks $T_{\mathrm{bc\overline{b}%
\overline{c}}}$ can easily decay to $D^{+}D^{-}$, $D^{0}\overline{D}^{0}$, $%
D^{\ast }{}^{+}D^{\ast -}$, and $D^{\ast }{}^{0}\overline{D}^{\ast }{}^{0}$
mesons: These processes are similar to ones considered in Refs.\ \cite%
{Becchi:2020mjz, Becchi:2020uvq,Agaev:2023ara} in the case of structures $bb%
\overline{b}\overline{b}/cc\overline{c}\overline{c} $. We calculate the full
width of the tetraquark $T_{\mathrm{bc\overline{b}\overline{c}}}$ by taking
into account these eight decay channels. Because partial widths of
aforementioned modes are determined by strong couplings at corresponding
tetraquark-meson-meson vertices, we estimate them by employing the
three-point SR method. The sum rule computations actually give information
about the strong form factor for a vertex under consideration, but these
data can be employed to find an extrapolating function and fix a coupling of
interest. In the case of the fall-apart processes application of the SR
method is standard and straightforward. But to analyze decays of $T_{\mathrm{%
bc\overline{b}\overline{c}}}$ generated by constituent $b$-quarks
annihilation, we perform an intermediate step and replace $\langle \overline{%
b}b\rangle $ vacuum matrix element in the three-point correlation function
by the gluon condensate $\langle \alpha _{s}G^{2}/\pi \rangle $. There are,
in general, another channels for decays of the tetraquark $T_{\mathrm{bc%
\overline{b}\overline{c}}}$ to conventional particles. Transformations to $4$%
-leptons or $2$-leptons+quarkonium are among such processes. But in the
present work, we restrict ourselves by analysis of $T_{\mathrm{bc\overline{b}%
\overline{c}}}$ tetraquark's strong decays.

We organize this paper in the following manner: In Sec.\ \ref{sec:Scalar},
we evaluate the mass and current coupling of the tetraquark $T_{\mathrm{bc%
\overline{b}\overline{c}}}$. We consider the allowed fall-apart decays of $%
T_{\mathrm{bc\overline{b}\overline{c}}}$ in Sec.\ \ref{sec:ScalarWidths1}.
The partial widths of the channels $D^{+}D^{-}$, $D^{0}\overline{D}^{0}$, $%
D^{\ast }{}^{+}D^{\ast -}$, and $D^{\ast }{}^{0}\overline{D}^{\ast }{}^{0}$
are calculated in Sec.\ \ref{sec:ScalarWidths2}. Here, we find also the full
width of the scalar tetraquark $T_{\mathrm{bc\overline{b}\overline{c}}}$. In
Sec.\ \ref{sec:Conc}, we compare our prediction for the mass of $T_{\mathrm{%
bc\overline{b}\overline{c}}}$ with results available in the literature and
make our brief conclusions .


\section{The mass and current coupling of the scalar structure $T_{\mathrm{bc%
\overline{b}\overline{c}}}$}

\label{sec:Scalar}

The spectroscopic parameters $m$ and $\Lambda $ of the scalar
diquark-antidiquark state $T_{\mathrm{bc\overline{b}\overline{c}}}$ are
important quantities which characterize it as four-quark meson and determine
its allowed decay modes. These parameters can be evaluated using the
two-point sum rule method elaborated in Refs.\ \cite%
{Shifman:1978bx,Shifman:1978by}.

To determine the sum rules for the mass $m$ and current coupling $\Lambda $,
one needs to study the correlation function%
\begin{equation}
\Pi (p)=i\int d^{4}xe^{ipx}\langle 0|\mathcal{T}\{J(x)J^{\dag
}(0)\}|0\rangle ,  \label{eq:CF1}
\end{equation}%
where $J(x)$ is the interpolating current for the tetraquark $T_{\mathrm{bc%
\overline{b}\overline{c}}}$, and $\mathcal{T}$ indicates the time-ordering
product of two currents.

We model $T_{\mathrm{bc\overline{b}\overline{c}}}$ in the diquark-antidiquark picture as a state
composed of the scalar diquarks $b^{T}C\gamma _{5}c$ and $\overline{b}\gamma
_{5}C\overline{c}^{T}$ which are most stable two-quark structures \cite%
{Jaffe:2004ph}. Then the current $J(x)$ takes the following form
\begin{equation}
J(x)=[b_{a}^{T}(x)C\gamma _{5}c_{b}(x)][\overline{b}_{a}(x)\gamma _{5}C%
\overline{c}_{b}^{T}(x)],  \label{eq:CR1}
\end{equation}%
with $a$, $b$ being the color indices and $C$ is the charge conjugation
matrix. This current describes the scalar particle with the spin-parity $J^{%
\mathrm{P}}=0^{+}$.

The SRs for the mass and current coupling of the state $T_{\mathrm{bc%
\overline{b}\overline{c}}}$ can be extracted after computing the correlation
function $\Pi (p)$ using both the physical parameters of $T_{\mathrm{bc%
\overline{b}\overline{c}}}$ and quark-gluon degrees of freedom. The first
function $\Pi ^{\mathrm{Phys}}(p)$ forms the physical side of the sum rule,
whereas $\Pi ^{\mathrm{OPE}}(p)$ constitutes its QCD side.

The correlation function $\Pi ^{\mathrm{Phys}}(p)$ is given by the
expression
\begin{equation}
\Pi ^{\mathrm{Phys}}(p)=\frac{\langle 0|J|T_{\mathrm{bc\overline{b}\overline{%
c}}}\rangle \langle T_{\mathrm{bc\overline{b}\overline{c}}}|J^{\dagger
}|0\rangle }{m^{2}-p^{2}}+\cdots .  \label{eq:Phys1}
\end{equation}%
The term written down explicitly in Eq.\ (\ref{eq:Phys1}) is a contribution
to $\Pi ^{\mathrm{Phys}}(p)$ arising from the ground-level particle.
Contributions to the correlator coming from higher resonances and continuum
states are shown by dots.

To simplify $\Pi ^{\mathrm{Phys}}(p)$, we rewrite it in terms of the
parameters $m$ and $\Lambda $. For these purposes, we introduce the matrix
element
\begin{equation}
\langle 0|J|T_{\mathrm{bc\overline{b}\overline{c}}}\rangle =\Lambda .
\label{eq:ME1}
\end{equation}%
After simple manipulations, one gets
\begin{equation}
\Pi ^{\mathrm{Phys}}(p)=\frac{\Lambda ^{2}}{m^{2}-p^{2}}+\cdots .
\label{eq:Phys2}
\end{equation}%
The correlator $\Pi ^{\mathrm{Phys}}(p)$ contains a term which has the
trivial Lorentz structure proportional to $\mathrm{I}$, therefore $\Lambda
^{2}/(m^{2}-p^{2})$ in the right-hand side of Eq.\ (\ref{eq:Phys2}) is the
invariant amplitude $\Pi ^{\mathrm{Phys}}(p^{2})$ required for future
studies.

The function $\Pi (p)$ has to be also computed using the heavy quark
propagators and operator product expansion ($\mathrm{OPE}$) with certain
accuracy. This function forms the QCD side $\Pi ^{\mathrm{OPE}}(p)$ of the
SRs and consists of the perturbative contribution, as well as contains a
nonperturbative term which is proportional to $\langle \alpha _{s}G^{2}/\pi
\rangle $. This is connected with the fact that heavy quark propagators do
not contain light quark and mixed quark-gluon condensates. Therefore,
possible additional contributions to $\Pi ^{\mathrm{OPE}}(p)$ are
proportional to $\langle g_{s}^{3}G^{3}\rangle $ and $\langle \alpha
_{s}G^{2}/\pi \rangle ^{2}$, which we neglect in our following
investigations.

The $\Pi ^{\mathrm{OPE}}(p)$ expressed using the $b$ and $c$ quarks'
propagators reads
\begin{eqnarray}
&&\Pi ^{\mathrm{OPE}}(p)=i\int d^{4}xe^{ipx}\mathrm{Tr}\left[ \gamma _{5}%
\widetilde{S}_{c}^{b^{\prime }b}(-x)\gamma _{5}S_{b}^{a^{\prime }a}(-x)%
\right]  \notag \\
&&\times \mathrm{Tr}\left[ \gamma _{5}\widetilde{S}_{b}^{aa^{\prime
}}(x)\gamma _{5}S_{c}^{bb^{\prime }}(x)\right] ,  \label{eq:QCD1}
\end{eqnarray}%
where%
\begin{equation}
\widetilde{S}_{Q}(x)=CS_{Q}^{T}(x)C,\ Q=b,\ c.  \label{eq:Prop}
\end{equation}%
In Eq.\ (\ref{eq:QCD1}) $S_{b(c)}(x)$ are propagators of the $b$ and $c$%
-quarks \cite{Agaev:2020zad}.

The $\Pi ^{\mathrm{OPE}}(p)$ has also a simple Lorentz structure $\sim
\mathrm{I}$, and is characterized by the invariant amplitude $\Pi ^{\mathrm{%
OPE}}(p^{2})$. The SRs for the $m$ and $\Lambda $ are obtained by equating $%
\Pi ^{\mathrm{OPE}}(p^{2})$ and $\Pi ^{\mathrm{Phys}}(p^{2})$, applying the
Borel transformation and performing continuum subtraction in accordance with
the assumption on quark-hadron duality \cite{Shifman:1978bx,Shifman:1978by}.
After these operations, we find the SRs for the mass and current coupling
\begin{equation}
m^{2}=\frac{\Pi ^{\prime }(M^{2},s_{0})}{\Pi (M^{2},s_{0})},  \label{eq:Mass}
\end{equation}%
and
\begin{equation}
\Lambda ^{2}=e^{m^{2}/M^{2}}\Pi (M^{2},s_{0}).  \label{eq:Coupl}
\end{equation}%
Here $\Pi (M^{2},s_{0})$ is the amplitude $\Pi ^{\mathrm{OPE}}(p^{2})$ after
the Borel transformation and continuum subtraction. It is a function of the
Borel and continuum subtraction parameters $M^{2}$ and $s_{0}$. Above we
used also the short-hand notation $\Pi ^{\prime }(M^{2},s_{0})=d\Pi
(M^{2},s_{0})/d(-1/M^{2})$.

The amplitude $\Pi (M^{2},s_{0})$ is calculated as an integral of the
two-point spectral density $\rho ^{\mathrm{OPE}}(s)$
\begin{equation}
\Pi (M^{2},s_{0})=\int_{4\mathcal{M}^{2}}^{s_{0}}ds\rho ^{\mathrm{OPE}%
}(s)e^{-s/M^{2}},  \label{eq:InvAmp}
\end{equation}%
where $\mathcal{M=(}m_{b}+m_{c})$. The spectral density$\rho ^{\mathrm{OPE}%
}(s)$ is determined as an imaginary part of the invariant amplitude $\Pi ^{%
\mathrm{OPE}}(p^{2})$ and is a sum of perturbative $\rho ^{\mathrm{pert.}%
}(s) $ and nonperturbative $\rho ^{\mathrm{Dim4}}(s)$ terms, which are given
by the general expression%
\begin{equation}
\rho (s)=\int_{0}^{1}d\alpha \int_{0}^{1-\alpha }d\beta \int_{0}^{1-\alpha
-\beta }d\gamma \rho (s,\alpha ,\beta ,\gamma ),
\end{equation}%
with $\alpha $, $\beta $, and $\gamma $ being the Feynman parameters. The
function $\rho ^{\mathrm{pert.}}(s,\alpha ,\beta ,\gamma )$ has the
following form
\begin{eqnarray}
&&\rho ^{\mathrm{pert.}}(s,\alpha ,\beta ,\gamma )=\frac{3N^{2}\Theta ({N)}}{%
1024A^{4}B^{2}C^{4}\pi ^{6}}\left\{ 6B^{2}C^{2}\right.  \notag \\
&&\times (L_{1}s\alpha ^{2}\beta ^{2}\gamma
-A^{2}m_{b}m_{c})(A^{2}m_{b}m_{c}+L_{1}^{2}s\alpha \beta \gamma ^{2})  \notag
\\
&&-A^{4}N\left[ 12CL_{1}^{2}s\alpha ^{2}\beta ^{2}\gamma ^{2}+B^{2}\left(
-3L_{1}N\alpha \beta \gamma +4Cm_{b}m_{c}\right. \right.  \notag \\
&&\left. \left. \times (\alpha (\beta -\gamma )+\gamma -\gamma (\beta
+\gamma )))\right] \right\} ,
\end{eqnarray}%
where $\Theta ({z)}$ is the Unit Step function. Here%
\begin{equation}
N=-C\left[ s\alpha \beta \gamma L_{1}+D(m_{c}^{2}L_{2}-m_{b}^{2}(\alpha
+\beta ))\right] /D^{2},
\end{equation}%
and
\begin{eqnarray}
&&A=L_{4}C+\alpha ^{2}(\beta +\gamma ),\ B=L_{2}C+\gamma ^{2}(\beta +\alpha
),  \notag \\
&&C=\alpha \beta +\alpha \gamma +\beta \gamma ,\ D=\gamma \beta L_{4}+\alpha
^{2}(\beta +\gamma )  \notag \\
&&+\alpha (\beta ^{2}+\gamma (\gamma -1)+\beta (2\gamma -1)).
\end{eqnarray}%
We also use the notations

\begin{eqnarray}
L_{1} &=&\alpha +\beta +\gamma -1,\ L_{2}=\alpha +\beta -1,  \notag \\
L_{3} &=&\alpha +\gamma -1,\ L_{4}=\beta +\gamma -1.
\end{eqnarray}%
An explicit formula for $\rho ^{\mathrm{Dim4}}(s,\alpha ,\beta ,\gamma )$ is
rather cumbersome and is not provided here.

The SRs for the mass $m$ and current coupling $\Lambda $ depend on the
masses of the heavy quarks and on the gluon condensate%
\begin{eqnarray}
&&m_{b}=4.18_{-0.02}^{+0.03}~\mathrm{GeV}\text{,\ }m_{c}=(1.27\pm 0.02)~%
\mathrm{GeV}\text{,}  \notag \\
&&\langle \alpha _{s}G^{2}/\pi \rangle =(0.012\pm 0.004)~\mathrm{GeV}^{4}.
\label{eq:GluonCond}
\end{eqnarray}%
There are also in Eqs.\ (\ref{eq:Mass}) and (\ref{eq:Coupl}) the parameters $%
M^{2}$ and $s_{0}$ a choice of which must satisfy constraints typical for
the sum rule method. In other words, they should lead to prevalence of the
pole contribution ($\mathrm{PC}$)
\begin{equation}
\mathrm{PC}=\frac{\Pi (M^{2},s_{0})}{\Pi (M^{2},\infty )},  \label{eq:PC}
\end{equation}%
by meeting the restriction $\mathrm{PC}\geq 0.5$. The upper limit of $M^{2}$
is extracted from this constraint. The convergence of $\mathrm{OPE}$
determines the lower bound for the Borel parameter. In our investigation
there is a nonperturbative term with dimension $4$, therefore we fix a
minimal value of $M^{2}$ by restricting its contribution by $\pm (5-15)\%$
of $\Pi (M^{2},s_{0})$. This constraint also leads to dominance of the
perturbative contribution in $\Pi (M^{2},s_{0})$. Another important
constraint is a stability of obtained quantities on the parameter $M^{2}$.

\begin{figure}[h]
\includegraphics[width=8.5cm]{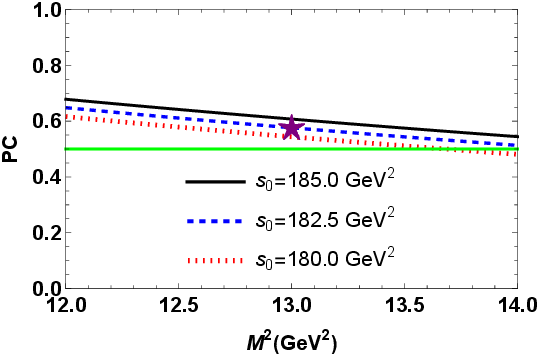}
\caption{The pole contribution $\mathrm{PC}$ as a function of $M^{2}$ at
fixed $s_{0}$. The red star marks the point $M^{2}=13~\mathrm{GeV}^{2}$ and $%
s_{0}=182.5~\mathrm{GeV}^{2}$. }
\label{fig:PC}
\end{figure}

Our numerical analysis demonstrates that these conditions are satisfied by
the working windows
\begin{equation}
M^{2}\in \lbrack 12,14]~\mathrm{GeV}^{2},\ s_{0}\in \lbrack 180,185]~\mathrm{%
GeV}^{2}.  \label{eq:Wind1}
\end{equation}%
Really, at $M^{2}=14~\mathrm{GeV}^{2}$ on the average in $s_{0}$ the pole
contribution is equal to $\mathrm{PC}\approx 0.51$, whereas at $M^{2}=12~%
\mathrm{GeV}^{2}$ it amounts to $\approx 0.65$. The nonperturbative
contribution is negative and at $M^{2}=12~\mathrm{GeV}^{2}$ constitutes only
$1\%$ of the whole result. In Fig.\ \ref{fig:PC}, we depict the pole
contribution $\mathrm{PC}$ as a function of the Borel parameter at the fixed
$s_{0}$. As is seen, it exceeds $0.5$ at all values of $M^{2}$ and $s_{0}$.

\begin{widetext}

\begin{figure}[h!]
\begin{center}
\includegraphics[totalheight=6cm,width=8cm]{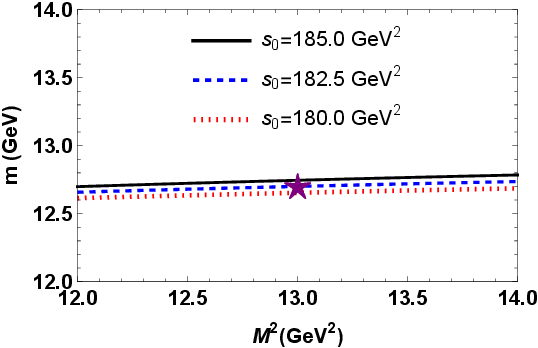}
\includegraphics[totalheight=6cm,width=8cm]{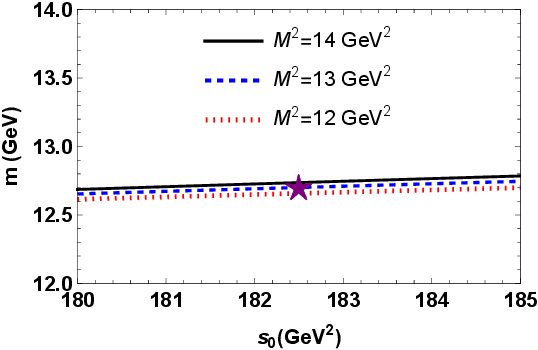}
\end{center}
\caption{Mass $m$ of the tetraquark $T_{\mathrm (2bc)}$ as a function of the Borel  $M^{2}$ (left panel), and continuum threshold $s_0$ parameters (right panel).}
\label{fig:Mass}
\end{figure}

\end{widetext}

The mass $m$ and current coupling $\Lambda $ are calculated as mean values
of these parameters over the regions Eq.\ (\ref{eq:Wind1}) and are equal to
\begin{eqnarray}
m &=&(12697\pm 90)~\mathrm{MeV},  \notag \\
\Lambda &=&(1.50\pm 0.17)~\mathrm{GeV}^{5}.  \label{eq:Result1}
\end{eqnarray}%
The predictions Eq.\ (\ref{eq:Result1}) effectively correspond to SR results
at the point $M^{2}=13~\mathrm{GeV}^{2}$ and $s_{0}=182.5~\mathrm{GeV}^{2}$,
where $\mathrm{PC}\approx 0.58$. This fact guarantees the dominance of $%
\mathrm{PC}$ in the extracted results, and demonstrates ground-level nature
of the tetraquark $T_{\mathrm{bc\overline{b}\overline{c}}}$ in its class.
Ambiguities in Eq.\ (\ref{eq:Result1}) are mainly generated by choices of
the parameters $M^{2}$ and $s_{0}$. In the case of the mass $m$ they form
only $\pm 0.07\%$ of the obtained result which implies very high stability
of performed analysis. This fact can be explained by the SR Eq.\ (\ref%
{eq:Mass}) defined as a ratio of two correlation functions. As a result,
variations in correlators due to $M^{2}$, $s_{0}$ and $m_{b}$,\ $m_{c}$ are
damped in $m$ which stabilizes numerical output. In the case of $\Lambda $
ambiguities are equal to $\pm 11\%$ remaining within acceptable limits of
the sum rule analysis. In Fig.\ \ref{fig:Mass}, we depict $m$ as a function
of $M^{2}$ and $s_{0}$.

\section{Decays of $T_{\mathrm{bc\overline{b}\overline{c}}}$: Fall-apart
processes}

\label{sec:ScalarWidths1}


The mass $m$ of the scalar tetraquark $T_{\mathrm{bc\overline{b}\overline{c}}%
}$ allows us to fix its kinematically allowed decay modes. It is clear that
a tetraquark with a content $bc\overline{b}\overline{c}$ may dissociate to
conventional mesons $\overline{c}b+\overline{b}c$ and $\overline{c}c+%
\overline{b}b$ with appropriate quantum numbers provided its mass exceeds
the mass of the final two-meson states. These decays are fall apart
processes in which constituent quarks are distributed between final-state
mesons.

Information on masses of the ordinary $\overline{c}b(\overline{b}c)$ and $%
\overline{c}c(\overline{b}b)$ mesons can be found in Ref.\ \cite{PDG:2022}.
It is worth noting that only the mass of the pseudoscalar ground-level and
radially excited mesons $B_{c}^{\pm }$ and $B_{c}^{\pm }(2S)$ are measured
experimentally. Parameters of particles with the same content but other
quantum numbers were calculated in the context of different methods. In the
present work, we employ predictions obtained in Ref.\ \cite{Godfrey:2004ya}.
The quarkonia $\overline{c}c(\overline{b}b)$ were investigated in detailed
form and their masses are known with a rather high precision. Corresponding
information about parameters of the mesons which will be considered in this
and next section is collected in Table\ \ref{tab:Param}.

It is not difficult to see that dissociation to pairs of $\eta _{b}\eta _{c}$%
, $B_{c}^{+}B_{c}^{-}$, $B_{c}^{\ast +}B_{c}^{\ast -}$ and $%
B_{c}^{+}(1^{3}P_{0})B_{c}^{\ast -}$ mesons are kinematically allowed decay
channels of the tetraquark $T_{\mathrm{bc\overline{b}\overline{c}}}$. In
this section, we study these modes and calculate their partial widths.


\subsection{$T_{\mathrm{bc\overline{b}\overline{c}}}\rightarrow \protect\eta %
_{b}\protect\eta _{c}$}


We begin from investigation of the decay $T_{\mathrm{bc\overline{b}\overline{%
c}}}\rightarrow \eta _{b}\eta _{c}$. The width of this process, apart from
parameters of particles involved into this decay, depends also on the strong
coupling $g_{1}$ at the vertex $T_{\mathrm{bc\overline{b}\overline{c}}}\eta
_{b}\eta _{c}$. To determine $g_{1}$, we study the QCD three-point
correlation function
\begin{eqnarray}
\Pi _{1}(p,p^{\prime }) &=&i^{2}\int d^{4}xd^{4}ye^{ip^{\prime
}y}e^{-ipx}\langle 0|\mathcal{T}\{J^{\eta _{b}}(y)  \notag \\
&&\times J^{\eta _{c}}(0)J^{\dagger }(x)\}|0\rangle ,  \label{eq:CF3}
\end{eqnarray}%
where
\begin{eqnarray}
J^{\eta _{b}}(x) &=&\overline{b}_{i}(x)i\gamma _{5}b_{i}(x),  \notag \\
J^{\eta _{c}}(x) &=&\overline{c}_{j}(x)i\gamma _{5}c_{j}(x),  \label{eq:CR3}
\end{eqnarray}%
are the interpolating currents of the pseudoscalar quarkonia $\eta _{b}$ and
$\eta _{c}$, respectively.

By studying the correlation function $\Pi _{1}(p,p^{\prime })$ it is
possible to derive the SR for the strong form factor $g_{1}(q^{2})$, which
at the mass shell $q^{2}=m_{\eta _{c}}^{2}$ becomes equal to $g_{1}$. To
determine the sum rule for the form factor $g_{1}(q^{2})$, we employ
well-known recipes of the method, and first write down $\Pi _{1}(p,p^{\prime
})$ in terms of masses and current coupling (decay constants) of the
tetraquark $T_{\mathrm{bc\overline{b}\overline{c}}}$ and quarkonia $\eta
_{b} $ and $\eta _{c}$. The correlation function $\Pi _{1}^{\mathrm{Phys}%
}(p,p^{\prime })$ calculated by this way forms the physical side of the sum
rule and is given by the expression
\begin{eqnarray}
&&\Pi _{1}^{\mathrm{Phys}}(p,p^{\prime })=\frac{\langle 0|J^{\eta _{b}}|\eta
_{b}(p^{\prime })\rangle }{p^{\prime 2}-m_{\eta _{b}}^{2}}\frac{\langle
0|J^{\eta _{c}}|\eta _{c}(q)\rangle }{q^{2}-m_{\eta _{c}}^{2}}  \notag \\
&&\times \langle \eta _{b}(p^{\prime })\eta _{c}(q)|T_{\mathrm{bc\overline{b}%
\overline{c}}}(p)\rangle \frac{\langle T_{\mathrm{bc\overline{b}\overline{c}}%
}(p)|J^{\dagger }|0\rangle }{p^{2}-m^{2}}+\cdots ,  \notag \\
&&  \label{eq:CF5}
\end{eqnarray}%
where $m_{\eta _{b}}$ and $m_{\eta _{c}}$ are the masses of $\eta _{b}$ and $%
\eta _{c}$. The correlation function $\Pi _{1}^{\mathrm{Phys}}(p,p^{\prime
}) $ is obtained after separating a contribution of the ground-level
particles from ones due to higher resonances and continuum states denoted
above by the dots.

We continue to simplify Eq.\ (\ref{eq:CF5}) by introducing the matrix
elements
\begin{eqnarray}
\langle 0|J^{\eta _{b}}|\eta _{b}(p^{\prime })\rangle &=&\frac{f_{\eta
_{b}}m_{\eta _{b}}^{2}}{2m_{b}},  \notag \\
\langle 0|J^{\eta _{c}}|\eta _{c}(q)\rangle &=&\frac{f_{\eta _{c}}m_{\eta
_{c}}^{2}}{2m_{c}},  \label{eq:ME2}
\end{eqnarray}%
with $f_{\eta _{b}}$ and $f_{\eta _{c}}$ being the decay constants of the
mesons $\eta _{b}$ and $\eta _{c}$ \cite{Veliev:2010vd}, respectively. We
model the vertex $T_{\mathrm{bc\overline{b}\overline{c}}}\eta _{b}\eta _{c}$
by means of the formula%
\begin{equation}
\langle \eta _{b}(p^{\prime })\eta _{c}(q)|T_{\mathrm{bc\overline{b}%
\overline{c}}}(p)\rangle =g_{1}(q^{2})p\cdot p^{\prime }.  \label{eq:ME3}
\end{equation}%
Then, the correlation function $\Pi _{1}^{\mathrm{Phys}}(p,p^{\prime })$
takes the following form
\begin{eqnarray}
&&\Pi _{1}^{\mathrm{Phys}}(p,p^{\prime })=g_{1}(q^{2})\frac{\Lambda f_{\eta
_{b}}m_{\eta _{b}}^{2}f_{\eta _{c}}m_{\eta _{c}}^{2}}{8m_{b}m_{c}\left(
p^{2}-m^{2}\right) }  \notag \\
&&\times \frac{m^{2}+m_{\eta _{b}}^{2}-q^{2}}{\left( p^{\prime 2}-m_{\eta
_{b}}^{2}\right) (q^{2}-m_{\eta _{c}}^{2})}+\cdots .  \label{eq:CF6}
\end{eqnarray}%
As is seen, $\Pi _{1}^{\mathrm{Phys}}(p,p^{\prime })$ has a simple Lorentz
structure proportional to $\mathrm{I}$, therefore we accept the right-hand
side of Eq.\ (\ref{eq:CF6}) as the invariant amplitude and denote it by $\Pi
_{1}^{\mathrm{Phys}}(p^{2},p^{\prime 2},q^{2})$.

\begin{table}[tbp]
\begin{tabular}{|c|c|}
\hline\hline
Quantity & Value (in $\mathrm{MeV}$ units) \\ \hline
$m_{\eta_{b}}$ & $9398.7\pm 2.0$ \\
$m_{\eta_c}$ & $2983.9\pm 0.4$ \\
$m_{B_{c}}$ & $6274.47\pm 0.27 \pm 0.17$ \\
$m_{B_{c}^{\ast}}$ & $6338$ \\
$m_{B_{c1}}$ & $6706$ \\
$m_{D}$ & $1869.5 \pm 0.05$ \\
$m_{D^0}$ & $1864.84 \pm 0.05$ \\
$m_{D^{\ast}}$ & $2010.26 \pm 0.05$ \\
$m_{D^{\ast 0}}$ & $2006.85 \pm 0.05$ \\
$f_{\eta_b}$ & $724$ \\
$f_{\eta_c}$ & $421 \pm 35$ \\
$f_{B_{c}}$ & $371 \pm 37$ \\
$f_{B_{c}^{\ast}}$ & $471$ \\
$f_{B_{c1}}$ & $236 \pm 17$ \\
$f_{D}$ & $211.9 \pm 1.1$ \\
$f_{D^{\ast}}$ & $252.2 \pm 22.66$ \\ \hline\hline
\end{tabular}%
\caption{Physical parameters of the conventional mesons, which have been
employed in numerical computations.}
\label{tab:Param}
\end{table}

The same correlator $\Pi _{1}(p,p^{\prime })$ computed in terms of the heavy
quark propagators is
\begin{eqnarray}
&&\Pi _{1}^{\mathrm{OPE}}(p,p^{\prime })=i^{4}\int d^{4}xd^{4}ye^{ip^{\prime
}y}e^{-ipx}\mathrm{Tr}\left[ \gamma _{5}\widetilde{S}_{b}^{ia}(y-x)\right.
\notag \\
&&\left. \times \gamma _{5}\widetilde{S}_{b}^{ai}(x-y)\gamma
_{5}S_{c}^{bj}(x)\gamma _{5}S_{c}^{jb}(-x)\right] .  \label{eq:QCDside2}
\end{eqnarray}%
We stand the function $\Pi _{1}^{\mathrm{OPE}}(p^{2},p^{\prime 2},q^{2})$
for the invariant amplitude that correspond in $\Pi _{1}^{\mathrm{OPE}%
}(p,p^{\prime })$ to term $\sim \mathrm{I}$, and employ it to find the SR
for the form factor $g_{1}(q^{2})$%
\begin{eqnarray}
&&g_{1}(q^{2})=\frac{8m_{b}m_{c}}{\Lambda f_{\eta _{b}}m_{\eta
_{b}}^{2}f_{\eta _{c}}m_{\eta _{c}}^{2}}\frac{q^{2}-m_{\eta _{c}}^{2}}{%
m^{2}+m_{\eta _{b}}^{2}-q^{2}}  \notag \\
&&\times e^{m^{2}/M_{1}^{2}}e^{m_{\eta _{b}}^{2}/M_{2}^{2}}\Pi _{1}(\mathbf{M%
}^{2},\mathbf{s}_{0},q^{2}),  \label{eq:SRCoup2}
\end{eqnarray}%
where
\begin{eqnarray}
&&\Pi _{1}(\mathbf{M}^{2},\mathbf{s}_{0},q^{2})=\int_{4\mathcal{M}%
^{2}}^{s_{0}}ds\int_{4m_{b}^{2}}^{s_{0}^{\prime }}ds^{\prime }\rho
_{1}(s,s^{\prime },q^{2})  \notag \\
&&\times e^{-s/M_{1}^{2}}e^{-s^{\prime }/M_{2}^{2}}.
\end{eqnarray}%
is the function $\Pi _{1}^{\mathrm{OPE}}(p^{2},p^{\prime 2},q^{2})$ after
the double Borel transformations over the variables $-p^{2}$, $-p^{\prime 2}$
and corresponding continuum subtractions procedures. It is written using the
spectral density $\rho _{1}(s,s^{\prime },q^{2})$.

The sum rule in Eq.\ (\ref{eq:SRCoup2}) depends on the Borel $\mathbf{M}%
^{2}=(M_{1}^{2},M_{2}^{2})$ and continuum threshold parameters $\mathbf{s}%
_{0}=(s_{0},s_{0}^{\prime })$. To perform numerical analysis, in the $T_{%
\mathrm{2bc}}$ channel we employ $M_{1}^{2}$ and $s_{0}$ given in Eq.\ (\ref%
{eq:Wind1}). The parameters $(M_{2}^{2},\ s_{0}^{\prime })$ in the $\eta
_{b} $ channel are varied inside of the borders%
\begin{equation}
M_{2}^{2}\in \lbrack 9,11]~\mathrm{GeV}^{2},\ s_{0}^{\prime }\in \lbrack
95,99]~\mathrm{GeV}^{2}.  \label{eq:Wind3}
\end{equation}%
It should be noted that$\ \sqrt{s_{0}^{\prime }}$ is limited by the mass $%
9.999\ \mathrm{GeV}$ of the first radially excited $\eta _{b}(2S)$ meson.

The credible results for the form factor $g_{1}(q^{2})$ in the sum rule
framework can be obtained in the Euclidean region $q^{2}<0$. At the same
time, the coupling $g_{1}$ should be fixed at the mass shell $q^{2}=m_{\eta
_{c}}^{2}$. To solve this problem, it is convenient to introduce a variable $%
Q^{2}=-q^{2}$ and employ $g_{1}(Q^{2})$ for the new function. Then, we use
an extrapolating function $\mathcal{G}_{1}(Q^{2})$ which for $Q^{2}>0$ gives
values coinciding with the SR results, but can be extended to the region of
negative $Q^{2}<0$. To this end, we use the function $\mathcal{G}_{i}(Q^{2})$
\begin{equation}
\mathcal{G}_{i}(Q^{2})=\mathcal{G}_{i}^{0}\mathrm{\exp }\left[ a_{i}^{1}%
\frac{Q^{2}}{m^{2}}+a_{i}^{2}\left( \frac{Q^{2}}{m^{2}}\right) ^{2}\right] ,
\end{equation}%
where $\mathcal{G}_{i}^{0}$, $a_{i}^{1}$, and $a_{i}^{2}$ are fitting
parameters.

\begin{figure}[h]
\includegraphics[width=8.5cm]{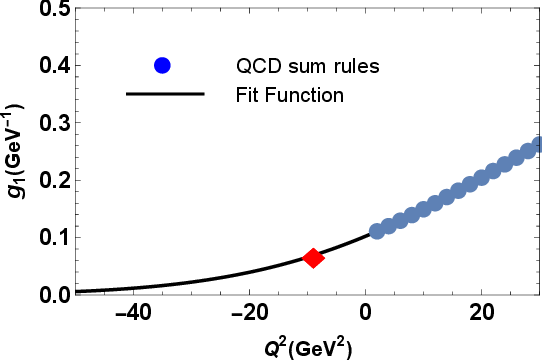}
\caption{QCD data and fit function for the form factor $g_{1}(Q^{2})$. The
diamond fix the point $Q^{2}=-m_{\protect\eta_c}^{2}$ where $g_{1}$ has been
evaluated. }
\label{fig:Fit}
\end{figure}

In this study, SR calculations cover the region $Q^{2}=2-30~\mathrm{GeV}^{2}$%
. Results of numerical calculations are shown in Fig.\ \ref{fig:Fit}. It is
not difficult to find that the function $\mathcal{G}_{1}(Q^{2})$ with
parameters $\mathcal{G}_{1}^{0}=0.10~\mathrm{GeV}^{-1}$, $a_{1}^{1}=7.03$,
and $a_{1}^{2}=-9.46$ leads to reasonable agreement with the SR data: This
function is also plotted in Fig.\ \ref{fig:Fit}.

As a result, we get for $g_{1}$
\begin{equation}
g_{1}\equiv \mathcal{G}_{1}(-m_{\eta _{c}}^{2})=(6.9\pm 0.9)\times 10^{-2}\
\mathrm{GeV}^{-1}.
\end{equation}%
The width of the decay $T_{\mathrm{bc\overline{b}\overline{c}}}\rightarrow
\eta _{b}\eta _{c}$ can be calculated by means of the expression%
\begin{equation}
\Gamma \left[ T_{\mathrm{bc\overline{b}\overline{c}}}\rightarrow \eta
_{b}\eta _{c}\right] =g_{1}^{2}\frac{m_{\eta _{b}}^{2}\lambda _{1}}{8\pi }%
\left( 1+\frac{\lambda _{1}^{2}}{m_{\eta _{b}}^{2}}\right) ,  \label{eq:PDw2}
\end{equation}%
where $\lambda _{1}=\lambda (m,m_{\eta _{b}},m_{\eta _{c}})$, and
\begin{equation}
\lambda (x,y,z)=\frac{\sqrt{%
x^{4}+y^{4}+z^{4}-2(x^{2}y^{2}+x^{2}z^{2}+y^{2}z^{2})}}{2x}.
\end{equation}%
Finally, we find
\begin{equation}
\Gamma \left[ T_{\mathrm{bc\overline{b}\overline{c}}}\rightarrow \eta
_{b}\eta _{c}\right] =(32.4\pm 8.3)~\mathrm{MeV}.  \label{eq:DW1}
\end{equation}

\subsection{$T_{\mathrm{bc\overline{b}\overline{c}}}\rightarrow
B_{c}^{+}B_{c}^{-}$}


The partial width of the decay $T_{\mathrm{bc\overline{b}\overline{c}}%
}\rightarrow B_{c}^{+}B_{c}^{-}$ is determined by the strong coupling $g_{2}$
at the vertex $T_{\mathrm{bc\overline{b}\overline{c}}}B_{c}^{+}B_{c}^{-}$.
In the framework of the QCD SR method the form factor $g_{2}(q^{2})$ is
calculated using the three-point correlation function%
\begin{eqnarray}
&&\Pi _{2}(p,p^{\prime })=i^{2}\int d^{4}xd^{4}ye^{ip^{\prime
}y}e^{-ipx}\langle 0|\mathcal{T}\{J^{B_{c}^{+}}(y)  \notag \\
&&\times J^{B_{c}^{-}}(0)J^{\dagger }(x)\}|0\rangle ,  \label{eq:CF7}
\end{eqnarray}%
where $J^{B_{c}^{+}}(y)$ and $J^{B_{c}^{-}}(0)$ are the interpolating
currents for the mesons $B_{c}^{+}$ and $B_{c}^{-}$
\begin{eqnarray}
&&J^{B_{c}^{+}}(x)=\overline{b}_{i}(x)i\gamma _{5}c_{i}(x),  \notag \\
&&J^{B_{c}^{-}}(x)=\overline{c}_{j}(x)i\gamma _{5}b_{j}(x),  \label{eq:CR4}
\end{eqnarray}%
respectively.

To determine the physical side of the SR, we use the matrix elements
\begin{equation}
\langle 0|J^{B_{c}^{\pm }}|B_{c}^{\pm }\rangle =\frac{f_{B_{c}}m_{B_{c}}^{2}%
}{m_{b}+m_{c}},  \label{eq:ME2A}
\end{equation}%
and
\begin{equation}
\langle B_{c}^{+}(p^{\prime })B_{c}^{-}(q)|T_{\mathrm{bc\overline{b}%
\overline{c}}}\rangle =g_{2}(q^{2})p\cdot p^{\prime },  \label{eq:ME3A}
\end{equation}%
where $m_{B_{c}}$ and $f_{B_{c}}$ are the mass and decay constant of the
mesons $B_{c}^{\pm }$ \cite{PDG:2022,Wang:2024fwc}. Then in terms of these
parameters the correlator $\Pi _{2}(p,p^{\prime })$ takes the form%
\begin{eqnarray}
&&\Pi _{2}^{\mathrm{Phys}}(p,p^{\prime })=\frac{g_{2}(q^{2})\Lambda
f_{B_{c}}^{2}m_{B_{c}}^{4}}{(m_{b}+m_{c})^{2}\left( p^{2}-m^{2}\right)
\left( p^{\prime 2}-m_{B_{c}}^{2}\right) }  \notag \\
&&\times \frac{m^{2}+m_{B_{c}}^{2}-q^{2}}{2(q^{2}-m_{B_{c}}^{2})}+\cdots .
\label{eq:CR2A}
\end{eqnarray}

The QCD side of the sum rule for the coupling $g_{2}(q^{2})$ is given by the
expression%
\begin{eqnarray}
&&\Pi _{2}^{\mathrm{OPE}}(p,p^{\prime })=-i^{4}\int
d^{4}xd^{4}ye^{ip^{\prime }y}e^{-ipx}\mathrm{Tr}\left[ \gamma
_{5}S_{c}^{ib}(y-x)\right.  \notag \\
&&\left. \times \gamma _{5}\widetilde{S}_{b}^{ja}(-x){}\gamma _{5}\widetilde{%
S}_{c}^{bj}(x)\gamma _{5}S_{b}^{ai}(x-y)\right] .  \label{eq:QCDside}
\end{eqnarray}%
In this decay the correlators $\Pi _{2}^{\mathrm{Phys}}(p,p^{\prime })$ and $%
\Pi _{2}^{\mathrm{OPE}}(p,p^{\prime })$ have simple Lorentz organizations.
Having denoted by $\Pi _{2}^{\mathrm{Phys}}(p^{2},p^{\prime 2},q^{2})$ and $%
\Pi _{2}^{\mathrm{OPE}}(p^{2},p^{\prime 2},q^{2})$ relevant invariant
amplitudes, we derive the following sum rule

\begin{eqnarray}
&&g_{2}(q^{2})=\frac{2(m_{b}+m_{c})^{2}}{\Lambda f_{B_{c}}^{2}m_{B_{c}}^{4}}%
\frac{q^{2}-m_{B_{c}}^{2}}{m^{2}+m_{B_{c}}^{2}-q^{2}}  \notag \\
&&\times e^{m^{2}/M_{1}^{2}}e^{m_{B_{c}}^{2}/M_{2}^{2}}\Pi _{2}(\mathbf{M}%
^{2},\mathbf{s}_{0},q^{2}).  \label{eq:SRCoup}
\end{eqnarray}%
Here, $\Pi _{2}(\mathbf{M}^{2},\mathbf{s}_{0},q^{2})$ is the Borel
transformed and subtracted amplitude $\Pi _{2}^{\mathrm{OPE}%
}(p^{2},p^{\prime 2},q^{2})$%
\begin{eqnarray}
&&\Pi _{2}(\mathbf{M}^{2},\mathbf{s}_{0},q^{2})=\int_{4\mathcal{M}%
^{2}}^{s_{0}}ds\int_{\mathcal{M}^{2}}^{s_{0}^{\prime }}ds^{\prime }\rho
_{2}(s,s^{\prime },q^{2})  \notag \\
&&\times e^{-s/M_{1}^{2}}e^{-s^{\prime }/M_{2}^{2}}.
\end{eqnarray}

The remaining operations are the same as ones explained above. Therefore, we
present only the windows for $M_{2}^{2}$, and $s_{0}^{\prime }$ in the $%
B_{c}^{+}$ meson channel
\begin{equation}
M_{2}^{2}\in \lbrack 6.5,7.5]~\mathrm{GeV}^{2},\ s_{0}^{\prime }\in \lbrack
45,47]~\mathrm{GeV}^{2}.
\end{equation}%
The extrapolating function $\mathcal{G}_{2}(Q^{2})$ has the parameters: $%
\mathcal{G}_{2}^{0}=0.17~\mathrm{GeV}^{-1}$, $a_{2}^{1}=1.97$, and $%
a_{2}^{2}=-1.41$. Then, the strong coupling $g_{2}$ becomes equal to
\begin{equation}
g_{2}\equiv \mathcal{G}_{2}(-m_{B_{c}}^{2})=(1.0\pm 0.2)\times 10^{-1}\
\mathrm{GeV}^{-1}.
\end{equation}%
Having used the strong coupling $g_{2}$, and Eq.\ (\ref{eq:PDw2}) with
evident replacements $m_{\eta _{b}}\rightarrow m_{B_{c}}$ and $\lambda
_{1}\rightarrow \lambda _{2}=\lambda (m,m_{B_{c}},m_{B_{c}})$, we find
\begin{equation}
\Gamma \left[ T_{\mathrm{bc\overline{b}\overline{c}}}\rightarrow B_{c}^{+}B_{c}^{-}\right]
=(33.9\pm 10.1)~\mathrm{MeV}.  \label{eq:DW2}
\end{equation}


\subsection{$T_{\mathrm{bc\overline{b}\overline{c}}}\rightarrow B_{c}^{\ast
+}B_{c}^{\ast -}$}


The next decay which will be considered in this subsection is the process $%
T_{\mathrm{bc\overline{b}\overline{c}}}\rightarrow B_{c}^{\ast +}B_{c}^{\ast
-}$. Treatment of this channel differs from previous modes by only some
technical details. Thus the correlation function necessary to be analyzed is
\begin{eqnarray}
\Pi _{\mu \nu }(p,p^{\prime }) &=&i^{2}\int d^{4}xd^{4}ye^{ip^{\prime
}y}e^{-ipx}\langle 0|\mathcal{T}\{J_{\mu }^{B_{c}^{\ast +}}(y)  \notag \\
&&\times J_{\nu }^{B_{c}^{\ast -}}(0)J^{\dagger }(x)\}|0\rangle .
\end{eqnarray}%
Here, $J_{\mu }^{B_{c}^{\ast +}}(y)$ and $J_{\nu }^{B_{c}^{\ast -}}(0)$ are
the interpolating currents of the vector mesons $B_{c}^{\ast +}$ and $%
B_{c}^{\ast -}$, respectively. They are given by the formulas%
\begin{eqnarray}
J_{\mu }^{B_{c}^{\ast +}}(x) &=&\overline{b}_{i}(x)\gamma _{\mu }c_{i}(x),
\notag \\
J_{\nu }^{B_{c}^{\ast -}}(x) &=&\overline{c}_{j}(x)\gamma _{\nu }b_{j}(x).
\end{eqnarray}

The physical side of the SR for the coupling $g_{3}(q^{2})$ which
corresponds to the strong vertex $T_{\mathrm{bc\overline{b}\overline{c}}}B_{c}^{\ast +}B_{c}^{\ast
-}$ is%
\begin{eqnarray}
&&\Pi _{\mu \nu }^{\mathrm{Phys}}(p,p^{\prime })=\frac{\langle 0|J_{\mu
}^{B_{c}^{\ast +}}|B_{c}^{\ast +}(p^{\prime })\rangle }{p^{\prime
2}-m_{B_{c}^{\ast }}^{2}}\frac{\langle 0|J_{\nu }^{B_{c}^{\ast
-}}|B_{c}^{\ast -}(q)\rangle }{q^{2}-m_{B_{c}^{\ast }}^{2}}  \notag \\
&&\times \langle B_{c}^{\ast +}(p^{\prime })B_{c}^{\ast -}(q)|T_{\mathrm{bc%
\overline{b}\overline{c}}}(p)\rangle \frac{\langle T_{\mathrm{bc\overline{b}%
\overline{c}}}(p)|J^{\dagger }|0\rangle }{p^{2}-m^{2}}+\cdots ,  \notag \\
&&
\end{eqnarray}%
where $m_{B_{c}^{\ast }}$ is the mass of the mesons $B_{c}^{\ast \pm }$. For
following studies it is convenient to introduce the matrix elements
\begin{eqnarray}
&&\langle 0|J_{\mu }^{B_{c}^{\ast +}}|B_{c}^{\ast +}(p^{\prime })\rangle
=f_{B_{c}^{\ast }}m_{B_{c}^{\ast }}\varepsilon _{\mu }(p^{\prime })  \notag
\\
&&\langle 0|J_{\nu }^{B_{c}^{\ast +}}|B_{c}^{\ast -}(q)\rangle
=f_{B_{c}^{\ast }}m_{B_{c}^{\ast }}\varepsilon _{\nu }(q),
\end{eqnarray}%
with $f_{B_{c}^{\ast }}$ and $\varepsilon (p^{\prime })$, $\varepsilon (q)$
being the decay constant and polarization vectors of $B_{c}^{\ast +}$ and $%
B_{c}^{\ast -}$. We model the vertex $T_{\mathrm{bc\overline{b}\overline{c}}%
}B_{c}^{\ast +}B_{c}^{\ast -}$ in the following form%
\begin{eqnarray}
&&\langle B_{c}^{\ast +}(p^{\prime })B_{c}^{\ast -}(q)|T_{\mathrm{bc%
\overline{b}\overline{c}}}(p)\rangle =g_{3}(q^{2})\left[ q\cdot p^{\prime
}\right.  \notag \\
&&\left. \times \varepsilon ^{\ast }(p^{\prime })\cdot \varepsilon ^{\ast
}(q)-q\cdot \varepsilon ^{\ast }(p^{\prime })p^{\prime }\cdot \varepsilon
^{\ast }(q)\right] .
\end{eqnarray}%
Then the $\Pi _{\mu \nu }(p,p^{\prime })$ in terms of physical parameters of
involved particles has the expression
\begin{eqnarray}
&&\Pi _{\mu \nu }^{\mathrm{Phys}}(p,p^{\prime })=\frac{g_{3}(q^{2})\Lambda
f_{B_{c}^{\ast }}^{2}m_{B_{c}^{\ast }}^{2}}{\left( p^{2}-m^{2}\right) \left(
p^{\prime 2}-m_{B_{c}^{\ast }}^{2}\right) (q^{2}-m_{B_{c}^{\ast }}^{2})}
\notag \\
&&\times \left[ \frac{1}{2}\left( m^{2}-m_{B_{c}^{\ast }}^{2}-q^{2}\right)
g_{\mu \nu }-q_{\mu }p_{\nu }^{\prime }\right] +\cdots .
\end{eqnarray}%
The correlator $\Pi _{\mu \nu }(p,p^{\prime })$ obtained using the heavy
quark propagators reads%
\begin{eqnarray}
&&\Pi _{\mu \nu }^{\mathrm{OPE}}(p,p^{\prime })=i^{2}\int
d^{4}xd^{4}ye^{ip^{\prime }y}e^{-ipx}\mathrm{Tr}\left[ \gamma _{\mu
}S_{c}^{ib}(y-x)\right.  \notag \\
&&\left. \times \gamma _{5}\widetilde{S}_{b}^{ja}(-x){}\gamma _{\nu }%
\widetilde{S}_{c}^{bj}(x)\gamma _{5}S_{b}^{ai}(x-y)\right] .
\end{eqnarray}%
The correlator $\Pi _{\mu \nu }(p,p^{\prime })$ in both cases contains the
two Lorentz structures $g_{\mu \nu }$ and $q_{\mu }p_{\nu }^{\prime }$. We
use invariant amplitudes that correspond to structures proportional to $%
g_{\mu \nu }$ and derive a required sum rule.

Numerical analysis has been carried out using the following input
parameters: For the Borel and continuum subtraction parameters in the $%
B_{c}^{\ast +}$ channel we have used
\begin{equation}
M_{2}^{2}\in \lbrack 6.5,7.5]~\mathrm{GeV}^{2},\ s_{0}^{\prime }\in \lbrack
50,51]~\mathrm{GeV}^{2}.
\end{equation}%
The decay constant $f_{B_{c}^{\ast }}$ of the mesons $B_{c}^{\ast }$ has
been chosen equal to $471~\mathrm{MeV}$ \cite{Eichten:2019gig}. The
extrapolating function $\mathcal{G}_{3}(Q^{2})$ has the parameters: $%
\mathcal{G}_{3}^{0}=0.22~\mathrm{GeV}^{-1}$, $a_{3}^{1}=2.65$, and $%
a_{3}^{2}=-3.58$. Then, the strong coupling $g_{3}$ amounts to
\begin{equation}
|g_{3}|\equiv \mathcal{G}_{3}(-m_{B_{c}^{\ast }}^{2})=(9.8\pm 1.3)\times
10^{-2}\ \mathrm{GeV}^{-1}.
\end{equation}%
The partial width of the decay $T_{\mathrm{bc\overline{b}\overline{c}}%
}\rightarrow B_{c}^{\ast +}B_{c}^{\ast -}$ can be found by means of the
formula
\begin{eqnarray}
\Gamma \left[ T_{\mathrm{bc\overline{b}\overline{c}}}\rightarrow B_{c}^{\ast
+}B_{c}^{\ast -}\right] &=&\frac{g_{3}^{2}\lambda _{3}}{16\pi }\left(
m^{2}-4m_{B_{c}^{\ast }}^{2}+6\frac{m_{B_{c}^{\ast }}^{4}}{m^{2}}\right) ,
\notag \\
&&
\end{eqnarray}%
where $\lambda _{3}=\lambda (m,m_{B_{c}^{\ast }},m_{B_{c}^{\ast }})$.
Numerical computations of the partial decay width of this channel give

\begin{equation}
\Gamma \left[ T_{\mathrm{bc\overline{b}\overline{c}}}\rightarrow B_{c}^{\ast
+}B_{c}^{\ast -}\right] =(21.1\pm 5.8)~\mathrm{MeV}.
\end{equation}


\subsection{$T_{\mathrm{bc\overline{b}\overline{c}}}\rightarrow
B_{c}^{+}(1^{3}P_{0})B_{c}^{\ast -}$}


Investigation of the decay $T_{\mathrm{bc\overline{b}\overline{c}}%
}\rightarrow B_{c}^{+}(1^{3}P_{0})B_{c}^{\ast -}$ has been performed in
accordance with the general scheme explained above. The correlation function
which should be considered in this case is given by the expression
\begin{eqnarray}
\Pi _{\mu }(p,p^{\prime }) &=&i^{2}\int d^{4}xd^{4}ye^{ip^{\prime
}y}e^{-ipx}\langle 0|\mathcal{T}\{J^{B_{c1}}(y)  \notag \\
&&\times J_{\mu }^{B_{c}^{\ast -}}(0)J^{\dagger }(x)\}|0\rangle ,
\end{eqnarray}%
with $J^{B_{c1}}(y)$ being the interpolating current of the scalar meson
$B_{c}^{+}(1^{3}P_{0})$
\begin{equation}
J^{B_{c1}}(y)=\overline{b}_{i}(x)c_{i}(x).
\end{equation}

The correlator $\Pi _{\mu }(p,p^{\prime })$ in terms of the particles'
parameters takes the form
\begin{eqnarray}
&&\Pi _{\mu }^{\mathrm{Phys}}(p,p^{\prime })=\frac{g_{4}(q^{2})\Lambda
f_{B_{c}^{\ast }}m_{B_{c}^{\ast }}f_{B_{c1}}m_{B_{c1}}}{\left(
p^{2}-m^{2}\right) \left( p^{\prime 2}-m_{B_{c1}}^{2}\right)
(q^{2}-m_{B_{c}^{\ast }}^{2})}  \notag \\
&&\times \left[ \frac{m^{2}-m_{B_{c1}}^{2}-q^{2}}{2m_{B_{c}^{\ast }}^{2}}%
p_{\mu }-\frac{m^{2}-m_{B_{c1}}^{2}+q^{2}}{2m_{B_{c}^{\ast }}^{2}}p_{\mu
}^{\prime }\right] +\cdots ,  \notag \\
&&
\end{eqnarray}%
where $m_{B_{c1}}$ and $f_{B_{c1}}$ are the mass and decay constant of $%
B_{c}^{+}(1^{3}P_{0})$ \cite{Godfrey:2004ya,Wang:2024fwc}. The function $\Pi
_{\mu }^{\mathrm{Phys}}(p,p^{\prime })$\ is obtained by using the new matrix
elements
\begin{equation}
\langle 0|J^{B_{c1}}|B_{c}^{+}(1^{3}P_{0})(p^{\prime })\rangle
=m_{B_{c1}}f_{B_{c1}},
\end{equation}%
and
\begin{equation}
\langle B_{c}^{+}(1^{3}P_{0})(p^{\prime })B_{c}^{\ast -}(q)|T_{\mathrm{bc%
\overline{b}\overline{c}}}(p)\rangle =g_{4}(q^{2})\varepsilon ^{\ast
}(q)\cdot p.
\end{equation}

The correlation function $\Pi _{\mu }(p,p^{\prime })$ calculated using the
heavy quark propagators is
\begin{eqnarray}
&&\Pi _{\mu }^{\mathrm{OPE}}(p,p^{\prime })=i^{2}\int
d^{4}xd^{4}ye^{ip^{\prime }y}e^{-ipx}\mathrm{Tr}\left[ S_{c}^{ib}(y-x)\right.
\notag \\
&&\left. \times \gamma _{5}\widetilde{S}_{b}^{ja}(-x){}\gamma _{\mu }%
\widetilde{S}_{c}^{bj}(x)\gamma _{5}S_{b}^{ai}(x-y)\right] .
\end{eqnarray}

Having omitted further details, we present final results for the parameters
of this process:%
\begin{equation}
|g_{4}|\equiv \mathcal{G}_{4}(-m_{B_{c}^{\ast }}^{2})=(12.2\pm 2.5),
\end{equation}%
and
\begin{equation}
\Gamma \left[ T_{\mathrm{bc\overline{b}\overline{c}}}\rightarrow
B_{c}^{+}(1^{3}P_{0})B_{c}^{\ast -}\right] =(25.4\pm 8.1)~\mathrm{MeV}.
\end{equation}


\section{Decays to charmed mesons}

\label{sec:ScalarWidths2}


This section is devoted to decays of $T_{\mathrm{bc\overline{b}\overline{c}}%
} $ to charmed mesons which become possible due to reorganization of this
particle's quark content. Thus, because $T_{\mathrm{bc\overline{b}\overline{c%
}}}$ contains a pair of $\overline{b}b$ (and $\overline{c}c)$ quarks, they
can annihilate and produce light quarks $q\overline{q}$ which later generate
conventional mesons. We are going to consider processes in which $T_{\mathrm{%
bc\overline{b}\overline{c}}}$ transforms to mesons $D^{+}D^{-}$, $D^{0}%
\overline{D}^{0}$, $D^{\ast +}D^{\ast -}$, and $D^{\ast 0}\overline{D}^{\ast
0}$. Effectively these channels are realized due to annihilation of $b%
\overline{b}$ quarks with subsequent creation of $u\overline{u}$ or $d%
\overline{d}$ pairs. In the context of the SR method these decays can be
considered using the three-point correlation functions and replacing the
heavy quark condensate $\langle \overline{b}b\rangle $ by the gluon
condensate $\langle \alpha _{s}G^{2}/\pi \rangle $.


\subsection{$T_{\mathrm{bc\overline{b}\overline{c}}}\rightarrow D^{0}%
\overline{D}^{0}$ and $D^{+}D^{-}$}


Let us analyze in a detailed form the process $T_{\mathrm{bc\overline{b}%
\overline{c}}}\rightarrow D^{0}\overline{D}^{0}$. In the case under
consideration the coupling $G_{1}$, which describes strong interaction at
the $T_{\mathrm{bc\overline{b}\overline{c}}}D^{0}\overline{D}^{0}$ vertex,
can be extracted from the three-point correlation function%
\begin{eqnarray}
\widetilde{\Pi }(p,p^{\prime }) &=&i^{2}\int d^{4}xd^{4}ye^{ip^{\prime
}y}e^{-ipx}\langle 0|\mathcal{T}\{J^{D^{0}}(y)  \notag \\
&&\times J^{\overline{D}^{0}}(0)J^{\dagger }(x)\}|0\rangle ,  \label{eq:CF1A}
\end{eqnarray}%
where the interpolating currents $J^{D^{0}}(y)$ and $J^{\overline{D}^{0}}(0)$
for the mesons $D^{0}=c\overline{u}$ and $\overline{D}^{0}=\overline{c}u$
have the following forms
\begin{equation}
J^{D^{0}}(x)=\overline{u}_{j}(x)i\gamma _{5}c_{j}(x),\ J^{\overline{D}%
^{0}}(x)=\overline{c}_{i}(x)i\gamma _{5}u_{i}(x).  \label{eq:CRB}
\end{equation}

\begin{figure}[h]
\includegraphics[width=8.5cm]{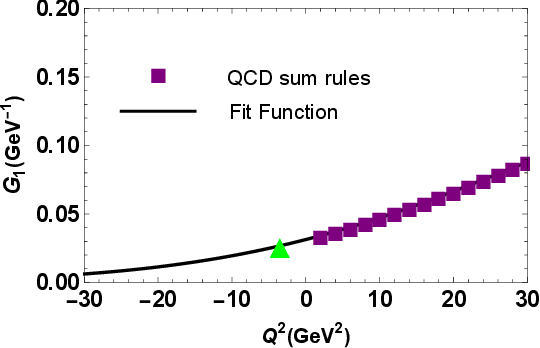}
\caption{Results of SR calculations and fit function for the form factor $%
G_{1}(Q^{2})$. The triangle shows the point $Q^{2}=-m_{D^0}^{2}$. }
\label{fig:Fit1}
\end{figure}

In accordance with the sum rule approach, we have to express the function $%
\Pi (p,p^{\prime })$ in terms of involved particles' parameters. By this
way, we determine the physical side of the sum rule. To this end, we write
down the $\Pi (p,p^{\prime })$ in the following form
\begin{eqnarray}
&&\widetilde{\Pi }^{\mathrm{Phys}}(p,p^{\prime })=\frac{\langle
0|J^{D^{0}}|D^{0}(p^{\prime })\rangle }{p^{\prime 2}-m_{D^{0}}^{2}}\frac{%
\langle 0|J^{\overline{D}^{0}}|\overline{D}^{0}(q)\rangle }{%
q^{2}-m_{D^{0}}^{2}}  \notag  \label{eq:CF2} \\
&&\times \langle D^{0}(p^{\prime })\overline{D}^{0}(q)|T_{\mathrm{bc%
\overline{b}\overline{c}}}(p)\rangle \frac{\langle T_{\mathrm{bc\overline{b}%
\overline{c}}}(p)|J^{\dagger }|0\rangle }{p^{2}-m^{2}}+\cdots ,  \notag \\
&&
\end{eqnarray}%
where $m_{D^{0}}$ is the $D^{0}$ and $\overline{D}^{0}$ mesons' mass.

To simplify $\widetilde{\Pi }^{\mathrm{Phys}}(p,p^{\prime })$, we express
the matrix elements in Eq.\ (\ref{eq:CF2}), using the masses and current
couplings (decay constants) of particles. The matrix element of the
pseudoscalar $D$ mesons is determined by the formula
\begin{equation}
\langle 0|J^{D^{0}}|D^{0}\rangle =\frac{f_{D}m_{D^{0}}^{2}}{m_{c}},
\label{eq:ME2B}
\end{equation}
with $f_{D}$ being the decay constant of $D^{0}$ and $\overline{D}^{0}$ \cite%
{Rosner:2015wva}. The matrix element of the meson $\overline{D}^{0}$ is
given by the same expression.

The vertex $\langle D^{0}(p^{\prime })\overline{D}^{0}(q)|T_{\mathrm{bc%
\overline{b}\overline{c}}}(p)\rangle $ is modeled by the formula
\begin{equation}
\langle D^{0}(p^{\prime })\overline{D}^{0}(q)|T_{\mathrm{bc\overline{b}%
\overline{c}}}(p)\rangle =G_{1}(q^{2})p\cdot p^{\prime }.  \label{eq:ME3B}
\end{equation}%
Here, $G_{1}(q^{2})$ is the strong form factor which at the mass shell of
the $\overline{D}^{0}$ meson, i.e., at $q^{2}=m_{D^{0}}^{2}$ fixes the
strong coupling $G_{1}$.

Then, it is easy to recast $\widetilde{\Pi }^{\mathrm{Phys}}(p,p^{\prime })$
into simpler form
\begin{eqnarray}
&&\widetilde{\Pi }^{\mathrm{Phys}}(p,p^{\prime })=G_{1}(q^{2})\frac{\Lambda
f_{D}^{2}m_{D}^{4}}{2m_{c}^{2}\left( p^{2}-m^{2}\right) \left( p^{\prime
2}-m_{D^{0}}^{2}\right) }  \notag \\
&&\times \frac{\left( m^{2}+m_{D^{0}}^{2}-q^{2}\right) }{%
(q^{2}-m_{D^{0}}^{2})}+\cdots ,  \label{eq:CorrF5}
\end{eqnarray}%
Because $\widetilde{\Pi }^{\mathrm{Phys}}(p,p^{\prime })$ has trivial
Lorentz organization, the whole expression in the right hand side of Eq.\ (%
\ref{eq:CorrF5}) is the invariant amplitude $\widetilde{\Pi }^{\mathrm{Phys}%
}(p^{2},p^{\prime 2},q^{2})$ which can be applied to derive the form factor $%
G_{1}(q^{2})$.

The first step in the current analysis does not differ from ones explained
in the previous section. The second component which is necessary to find SR
for $G_{1}(q^{2})$ is the correlator Eq.\ (\ref{eq:CF1A}) computed using the
quark propagators, which reads%
\begin{eqnarray}
&&\widetilde{\Pi }^{\mathrm{OPE}}(p,p^{\prime })=\int
d^{4}xd^{4}ye^{ip^{\prime }y}e^{-ipx}\langle \overline{b}b\rangle  \notag \\
&&\times \mathrm{Tr}\left[ \gamma _{5}S_{c}^{ja}(y-x){}S_{c}^{ai}(x-y)\gamma
_{5}S_{u}^{ij}(-y)\right] ,  \label{eq:QCDsideA}
\end{eqnarray}%
where $S_{u}(x)$ is the $u$ quark propagator \cite{Agaev:2020zad}.

The function $\widetilde{\Pi }^{\mathrm{OPE}}(p,p^{\prime })$ depends on
three quark propagators and vacuum condensate $\langle \overline{b}b\rangle $
of $b$ quarks. The function $\widetilde{\Pi }^{\mathrm{OPE}}(p,p^{\prime })$
differs from a standard correlator. Indeed, the latter in the case, for
example, of the decay$T_{\mathrm{bc\overline{b}\overline{c}}}\rightarrow
\eta _{b}\eta _{c}$ contains four propagators $S_{b(c)}(x)$. The reason is
that to calculate $\Pi ^{\mathrm{OPE}}(p,p^{\prime })$ we contract heavy and
light quark fields and, as a result, $\overline{b}b$ quarks establish a
heavy quark condensate $\langle \overline{b}b\rangle $, because \ mesons $%
D^{0}\overline{D}^{0}$ do not contain $b$ quarks.

Using the relation between different condensates
\begin{equation}
m_{b}\langle \overline{b}b\rangle =-\frac{1}{12\pi }\langle \frac{\alpha
_{s}G^{2}}{\pi }\rangle ,  \label{eq:Conden}
\end{equation}%
we get
\begin{eqnarray}
&&\widetilde{\Pi }^{\mathrm{OPE}}(p,p^{\prime })=-\frac{1}{12m_{b}\pi }%
\langle \frac{\alpha _{s}G^{2}}{\pi }\rangle \int d^{4}xd^{4}ye^{ip^{\prime
}y}e^{-ipx}  \notag \\
&&\times \mathrm{Tr}\left[ \gamma _{5}S_{c}^{ja}(y-x){}S_{c}^{ai}(x-y)\gamma
_{5}S_{u}^{ij}(-y)\right] .  \label{eq:CF6A}
\end{eqnarray}%
In other words, the correlator $\widetilde{\Pi }^{\mathrm{OPE}}(p,p^{\prime
})$ is suppressed by the factor $\langle \alpha _{s}G^{2}/\pi \rangle $. In
what follows, we denote the corresponding invariant amplitude as $\widetilde{%
\Pi }^{\mathrm{OPE}}(p^{2},p^{\prime 2},q^{2})$.

In calculation of $\Pi ^{\mathrm{OPE}}(p,p^{\prime })$, we set $m_{u}=0$.
The perturbative terms in all propagators lead to a contribution which is
proportional to $\langle \alpha _{s}G^{2}/\pi \rangle $. A dimension-7 term
in $\Pi ^{\mathrm{OPE}}(p,p^{\prime })$ arises from the component $\sim
\langle \overline{u}u\rangle $ in $S_{u}(x)$ and perturbative ones in $%
S_{b}(x)$. A contribution $\sim \langle \alpha _{s}G^{2}/\pi \rangle ^{2}$
appearing due to $g_{s}G_{ab}^{\alpha \beta }$ components in $b$ propagators
and perturbative term in $S_{u}(x)$ is small and can be neglected. Higher
dimensional terms in the quark propagators lead to effects suppressed by
additional factors. It is seen that the strong coupling $G_{1}$ depends on
the gluon vacuum condensate $\langle \alpha _{s}G^{2}/\pi \rangle $, which
was fixed from consideration of the different hadronic processes \cite%
{Shifman:1978bx,Shifman:1978by,Ioffe:2005ym,Narison:2018dcr}. In our study
we use its value from Eq.\ (\ref{eq:GluonCond}) extracted in Refs.\ \cite%
{Shifman:1978bx,Shifman:1978by}.

Operations required to extract SR for the strong form factor using
correlator $\widetilde{\Pi }^{\mathrm{OPE}}(p,p^{\prime })$ do not differ
from ones explained in this article. Therefore, omitting details, we present
final results for the decay $T_{\mathrm{bc\overline{b}\overline{c}}%
}\rightarrow D^{0}\overline{D}^{0}$. Numerical computations of the strong
form factor $G_{1}(q^{2})$ is carried out using the following Borel and
continuum \ subtraction parameters in the $D^{0}$ channel: $M_{2}^{2}\in
\lbrack 1.5,3]~\mathrm{GeV}^{2},\ s_{0}^{\prime }\in \lbrack 5,5.2]~\mathrm{%
GeV}^{2}$. The coupling $G_{1}$ is extracted at the mass shell $%
q^{2}=m_{D^{0}}^{2}$ and is equal to
\begin{equation}
G_{1}\equiv \widetilde{\mathcal{G}}_{1}(-m_{D^{0}}^{2})=(2.68\pm 0.42)\times
10^{-2}\ \mathrm{GeV}^{-1}.  \label{eq:G1}
\end{equation}%
To find $G_{1}$, we have employed the SR data calculated in the interval $%
Q^{2}=2-30\ \mathrm{GeV}^{2}$ and the fit function with parameters $%
\widetilde{\mathcal{G}}_{1}^{0}=0.031~\mathrm{GeV}^{-1}$, $\widetilde{a}%
_{1}^{1}=7.58$, and $\widetilde{a}_{1}^{2}=-9.61$. The SR data and
extrapolating function $\widetilde{\mathcal{G}}_{1}(Q^{2})$ are shown in
Fig.\ \ref{fig:Fit1}.

The with of this process can be obtained by means of the expression%
\begin{equation}
\Gamma \left[ T_{\mathrm{bc\overline{b}\overline{c}}}\rightarrow D^{0}%
\overline{D}^{0}\right] =G_{1}^{2}\frac{m_{D^{0}}^{2}\widetilde{\lambda }_{1}%
}{8\pi }\left( 1+\frac{\widetilde{\lambda }_{1}^{2}}{m_{D^{0}}^{2}}\right) ,
\end{equation}%
where $\widetilde{\lambda }_{1}=\lambda (m,m_{D^{0}}^{2},m_{D^{0}}^{2})$.
Finally, we get
\begin{equation}
\Gamma \left[ T_{\mathrm{bc\overline{b}\overline{c}}}\rightarrow D^{0}%
\overline{D}^{0}\right] =(7.6\pm 2.1)~\mathrm{MeV}.
\end{equation}

Parameters of the decay $T_{\mathrm{bc\overline{b}\overline{c}}}\rightarrow
D^{+}D^{-}$ should be very close to ones of the channel $T_{\mathrm{bc%
\overline{b}\overline{c}}}\rightarrow D^{0}\overline{D}^{0}$ a difference
being connected with small mass gap between the mesons $D^{0}$ and $D^{\pm }$%
. We neglect such unessential difference and set $\Gamma \left[ T_{\mathrm{bc%
\overline{b}\overline{c}}}\rightarrow D^{+}D^{-}\right] \approx \Gamma \left[
T_{\mathrm{bc\overline{b}\overline{c}}}\rightarrow D^{0}\overline{D}^{0}%
\right] .$


\subsection{$T_{\mathrm{bc\overline{b}\overline{c}}}\rightarrow D^{\ast 0}%
\overline{D}^{\ast 0}$ and $D^{\ast +}D^{\ast -}$}


Decays to vector charmed mesons is considered within the same scheme. Here,
we start form analysis of the correlation function
\begin{eqnarray}
\widetilde{\Pi }_{\mu \nu }(p,p^{\prime }) &=&i^{2}\int
d^{4}xd^{4}ye^{ip^{\prime }y}e^{-ipx}\langle 0|\mathcal{T}\{J_{\mu
}^{D^{\ast 0}}(y)  \notag \\
&&\times J_{\nu }^{\overline{D}^{\ast 0}}(0)J^{\dagger }(x)\}|0\rangle ,
\end{eqnarray}%
where the interpolating currents of the $D^{\ast 0}$ and $\overline{D}^{\ast
0}$ mesons are defined by means of the formulas%
\begin{eqnarray}
J_{\mu }^{D^{\ast 0}}(x) &=&\overline{u}_{i}(x)\gamma _{\mu }c_{i}(x),
\notag \\
J_{\nu }^{\overline{D}^{\ast 0}}(x) &=&\overline{c}_{j}(x)\gamma _{\nu
}u_{j}(x).
\end{eqnarray}%
The matrix elements which are necessary to calculate the correlator in terms
of masses and decay constants of the $D^{\ast }$ mesons are%
\begin{eqnarray}
&&\langle 0|J_{\mu }^{D^{\ast 0}}|D^{\ast 0}(p^{\prime })\rangle =f_{D^{\ast
}}m_{D^{\ast 0}}\varepsilon _{\mu }(p^{\prime })  \notag \\
&&\langle 0|J_{\nu }^{\overline{D}^{\ast 0}}|\overline{D}^{\ast 0}(q)\rangle
=f_{D^{\ast }}m_{D^{\ast 0}}\varepsilon _{\nu }(q),
\end{eqnarray}%
where $m_{D^{\ast 0}}$ and $f_{D^{\ast }}$ are the mass and decay constant
of $D^{\ast 0}$ and $\overline{D}^{\ast 0}$. The polarization vectors of
these particles are $\varepsilon _{\mu }(p^{\prime })$ and $\varepsilon
_{\nu }(q)$. We introduce the vertex by the following way

\begin{eqnarray}
&&\langle D^{\ast 0}(p^{\prime })\overline{D}^{\ast 0}(q)|T_{\mathrm{bc%
\overline{b}\overline{c}}}(p)\rangle =G_{2}(q^{2})\left[ q\cdot p^{\prime
}\right.  \notag \\
&&\left. \times \varepsilon ^{\ast }(p^{\prime })\cdot \varepsilon ^{\ast
}(q)-q\cdot \varepsilon ^{\ast }(p^{\prime })p^{\prime }\cdot \varepsilon
^{\ast }(q)\right] .
\end{eqnarray}%
Here, $G_{2}(q^{2})$ is the form factor which correspond to the vertex $T_{%
\mathrm{bc\overline{b}\overline{c}}}D^{\ast 0}\overline{D}^{\ast 0}$ and at
the mass shell $q^{2}=m_{D^{\ast 0}}^{2}$ is equal to the strong coupling $%
G_{2}$.

As a result, the $\widetilde{\Pi }_{\mu \nu }^{\mathrm{Phys}}(p,p^{\prime })$
takes the form%
\begin{eqnarray}
&&\widetilde{\Pi }_{\mu \nu }^{\mathrm{Phys}}(p,p^{\prime })=\frac{%
G_{2}(q^{2})\Lambda f_{D^{\ast }}^{2}m_{D^{\ast 0}}^{2}}{\left(
p^{2}-m^{2}\right) \left( p^{\prime 2}-m_{D^{\ast 0}}^{2}\right)
(q^{2}-m_{D^{\ast 0}}^{2})}  \notag \\
&&\times \left[ \frac{1}{2}\left( m^{2}-m_{D^{\ast 0}}^{2}-q^{2}\right)
g_{\mu \nu }-q_{\mu }p_{\nu }^{\prime }\right] +\cdots .
\end{eqnarray}%
The correlation function $\widetilde{\Pi }_{\mu \nu }(p,p^{\prime })$ in
terms of the quark propagators is equal to
\begin{eqnarray}
&&\widetilde{\Pi }_{\mu \nu }^{\mathrm{OPE}}(p,p^{\prime })=i^{2}\int
d^{4}xd^{4}ye^{ip^{\prime }y}e^{-ipx}\langle \overline{b}b\rangle  \notag \\
&&\times \mathrm{Tr}\left[ \gamma _{\mu }S_{c}^{ja}(y-x)\widetilde{S}%
_{c}^{aj}(x)\gamma _{\nu }S_{u}^{ij}(-y)\right] .
\end{eqnarray}

We use the invariant amplitudes which correspond to structures $g_{\mu \nu }$
in functions $\widetilde{\Pi }_{\mu \nu }^{\mathrm{Phys}}(p,p^{\prime })$
and $\widetilde{\Pi }_{\mu \nu }^{\mathrm{OPE}}(p,p^{\prime })$. In
numerical analysis the Borel and continuum subtraction parameters in the $%
D^{\ast 0}$ meson channel are chosen equal to $M_{2}^{2}\in \lbrack 2,3]~%
\mathrm{GeV}^{2},\ s_{0}^{\prime }\in \lbrack 5.7,5.8]~\mathrm{GeV}^{2}$.
The coupling $G_{2}$ is

\begin{equation}
G_{2}\equiv \widetilde{\mathcal{G}}_{2}(-m_{D^{\ast 0}}^{2})=(1.93\pm
0.33)\times 10^{-2}\ \mathrm{GeV}^{-1}.
\end{equation}%
Then, the partial width of the channel $T_{\mathrm{bc\overline{b}\overline{c}%
}}\rightarrow D^{\ast 0}\overline{D}^{\ast 0}$ amounts to
\begin{equation}
\Gamma \left[ T_{\mathrm{bc\overline{b}\overline{c}}}\rightarrow D^{\ast 0}%
\overline{D}^{\ast 0}\right] =(7.2\pm 1.9)~\mathrm{MeV}.
\end{equation}%
It is clear that an approximate equality $\Gamma \left[ T_{\mathrm{bc%
\overline{b}\overline{c}}}\rightarrow D^{\ast +}D^{\ast -}\right] \approx
\Gamma \left[ T_{\mathrm{bc\overline{b}\overline{c}}}\rightarrow D^{\ast 0}%
\overline{D}^{\ast 0}\right] $ is correct for these two decay modes as well.

Information collected in this and previous sections about partial widths of
different processes allows us to evaluate the full width of the scalar
tetraquark $bc\overline{b}\overline{c}$. For this parameter, we get
\begin{equation}
\Gamma \left[ T_{\mathrm{bc\overline{b}\overline{c}}}\right] =(142.4\pm
16.9)~\mathrm{MeV},
\end{equation}%
which classifies $T_{\mathrm{bc\overline{b}\overline{c}}}$ as a wide state.


\section{Discussion and concluding notes}

\label{sec:Conc}


In this paper, we have presented results of our comprehensive analysis of
the scalar four-quark meson with quark content $bc\overline{b}\overline{c}$.
We have considered it as a diquark-antidiquark state made of scalar diquark
and antidiquark components $b^{T}C\gamma _{5}c$ and $\overline{b}\gamma _{5}C%
\overline{c}^{T}$. We have calculated the mass and width of this structure
and found them equal to $m=(12697\pm 90)~\mathrm{MeV}$ and $\Gamma \left[ T_{%
\mathrm{bc\overline{b}\overline{c}}}\right] =(142.4\pm 16.9)~\mathrm{MeV}$,
respectively. Our studies have been carried out using the QCD SR methods.
Thus, the mass and current coupling of $T_{\mathrm{bc\overline{b}\overline{c}%
}}$ have been evaluated using the two-point SR approach.

Partial widths of different decay channels of the tetraquark $T_{\mathrm{bc%
\overline{b}\overline{c}}}$ have been computed by applying technique of
three-point SR method. This is necessary to estimate strong couplings at
relevant tetraquark-meson-meson vertices. We have analyzed two type of decay
processes. First, we have considered decays to $\eta_b\eta_c  $ and $B_{c}^{(\ast )}B_{c}^{(\ast
)}$ pairs which are fall-apart processes, where all constituent quarks are
distributed between final-state mesons. Second type of decays are ones which
become possible due to annihilation of $\overline{b}b$ to light quarks
followed by creation of a pair of $D^{(\ast )}D^{(\ast )}$ mesons with
appropriate charges and quantum numbers.

The masses of the tetraquarks $bc\overline{b}\overline{c}$ with different
quantum numbers $J^{\mathrm{PC}}$ were evaluated in various publications
\cite%
{Faustov:2022mvs,Wu:2016vtq,Liu:2019zuc,Chen:2019vrj,Bedolla:2019zwg,Cordillo:2020sgc,Weng:2020jao,Yang:2021zrc,Hoffer:2024alv}%
. In fact, in the relativistic quark model the mass of the scalar particle $%
bc\overline{b}\overline{c}$ was estimated $12813~\mathrm{MeV}$ \cite%
{Faustov:2022mvs}. The predictions of Ref.\ \cite{Wu:2016vtq}, where the
authors used the color-magnetic interaction, change in the range of $%
13396-13634~\mathrm{MeV}$. A potential model which includes the linear and
Coulomb potentials, and spin-spin interactions leads to the mass spectrum of
the scalar particle varying within limits $12854-13024~\mathrm{MeV}$ \cite%
{Liu:2019zuc}. In Ref.\ \cite{Yang:2021zrc} the authors applied the QCD SR
method and found that the mass of the diquark-antidiquark state with $J^{%
\mathrm{PC}}=0^{++}$ is equal to $12.28-12.46~\mathrm{GeV}$ depending on an
interpolating current used in computations. The lowest prediction $10.72~%
\mathrm{GeV}$ for the mass of the scalar state $bc\overline{b}\overline{c}$
was made recently in Ref.\ \cite{Hoffer:2024alv}.

Our result overshoots considerably the SR predictions made in Ref. \cite%
{Yang:2021zrc}, being close to one from \cite{Faustov:2022mvs} and to lower
limits found in some other publications. One of main reasons to explore the
structures $bc\overline{b}\overline{c}$ is intention to find a state stable
against strong decays. Therefore, in many articles authors compare obtained
results for the mass with various two-meson thresholds. But, as we have
demonstrated in previous section, without regard to output of such analysis,
four-quark mesons $bc\overline{b}\overline{c}$ are strong-interaction
unstable particles. Our estimates show that width of $T_{\mathrm{bc\overline{%
b}\overline{c}}}$ generated by annihilation subprocesses $\overline{b}%
b\rightarrow \overline{q}q$ is not small and form at least $20\%$ of its
full width.

Further detailed investigations of the tetraquarks $bc\overline{b}\overline{c%
}$ are necessary to calculate their parameters and find reactions where they
may be observed.


\begin{thebibliography}{99}


\bibitem{LHCb:2020bwg} R.~Aaij \textit{et al.} (LHCb Collaboration),
Sci.\ Bull. \textbf{65}, 1983 (2020).


\bibitem{Bouhova-Thacker:2022vnt} E.~Bouhova-Thacker (ATLAS Collaboration),
PoS \textbf{ICHEP2022}, 806 (2022).


\bibitem{CMS:2023owd} A.~Hayrapetyan, \textit{et al.} (CMS Collaboration)
Phys.\ Rev.\ Lett.\ \textbf{132}, 111901 (2024).


\bibitem{Zhang:2020xtb} J.~R.~Zhang,
Phys.\ Rev.\ D \textbf{103}, 014018 (2021).


\bibitem{Albuquerque:2020hio} R.~M.~Albuquerque, S.~Narison,
A.~Rabemananjara, D.~Rabetiarivony, and G.~Randriamanatrika,
Phys.\ Rev.\ D \textbf{102}, 094001 (2020).


\bibitem{Yang:2020wkh} B.~C.~Yang, L.~Tang, and C.~F.~Qiao,
Eur.\ Phys.\ J. C \textbf{81}, 324 (2021). 


\bibitem{Becchi:2020mjz} C.~Becchi, A.~Giachino, L.~Maiani, and
E.~Santopinto,
Phys.\ Lett.\ B \textbf{806}, 135495 (2020).


\bibitem{Becchi:2020uvq} C.~Becchi, A.~Giachino, L.~Maiani, and
E.~Santopinto, 
Phys.\ Lett.\ B \textbf{811}, 135952 (2020). 


\bibitem{Wang:2022xja} Z.~G.~Wang,
Nucl.\ Phys.\ B \textbf{985}, 115983 (2022). 


\bibitem{Faustov:2022mvs} R.~N.~Faustov, V.~O.~Galkin, and E.~M.~Savchenko,
Symmetry \textbf{14}, 2504 (2022). 


\bibitem{Niu:2022vqp} P.~Niu, Z.~Zhang, Q.~Wang, and M.~L.~Du,
Sci.\ Bull.\ \textbf{68}, 800 (2023). 


\bibitem{Dong:2022sef} W.~C.~Dong and Z.~G.~Wang,
Phys.\ Rev.\ D \textbf{107}, 074010 (2023). 


\bibitem{Yu:2022lak} G.~L.~Yu, Z.~Y.~Li, Z.~G.~Wang, J.~Lu, and M.~Yan,
Eur.\ Phys.\ J. C \textbf{83}, 416 (2023). 


\bibitem{Kuang:2023vac} S.~Q.~Kuang, Q.~Zhou, D.~Guo, Q.~H.~Yang, and
L.~Y.~Dai,
Eur.\ Phys.\ J. C \textbf{83}, 383 (2023). 


\bibitem{Wang:2023kir} Z.~G.~Wang and X.~S.~Yang,
AAPPS\ Bull.\ \textbf{34}, 5 (2024). 


\bibitem{Dong:2020nwy} X.~K.~Dong, V.~Baru, F.~K.~Guo, C.~Hanhart, and
A.~Nefediev,
Phys.\ Rev.\ Lett. \textbf{126}, 132001 (2021); \textbf{127}, 119901(E)
(2021). 


\bibitem{Liang:2021fzr} Z.~R.~Liang, X.~Y.~Wu, and D.~L.~Yao,
Phys.\ Rev.\ D \textbf{104}, 034034 (2021).


\bibitem{Agaev:2023wua} S.~S.~Agaev, K.~Azizi, B.~Barsbay, and H.~Sundu,
Phys.\ Lett.\ B \textbf{844}, 138089 (2023). 


\bibitem{Agaev:2023ruu} S.~S.~Agaev, K.~Azizi, B.~Barsbay and H.~Sundu,
Eur.\ Phys.\ J. Plus \textbf{138}, 935 (2023). 


\bibitem{Agaev:2023gaq} S.~S.~Agaev, K.~Azizi, B.~Barsbay and H.~Sundu,
Nucl.\ Phys.\ A \textbf{844}, 122768 (2024). 


\bibitem{Agaev:2023rpj} S.~S.~Agaev, K.~Azizi, B.~Barsbay and H.~Sundu,
Eur.\ Phys.\ J. C \textbf{83}, 994 (2023). 


\bibitem{Wu:2016vtq} J.~Wu, Y.~R.~Liu, K.~Chen, X.~Liu, and S.~L.~Zhu,
Phys.\ Rev.\ D \textbf{97}, 094015 (2018). 


\bibitem{Li:2019uch} G.~Li, X.~F.~Wang, and Y.~Xing,
Eur.\ Phys.\ J.\ C \textbf{79}, 645 (2019). 


\bibitem{Wang:2019rdo} G.~J.~Wang, L.~Meng, and S.~L.~Zhu,
Phys.\ Rev.\ D \textbf{100}, 096013 (2019). 


\bibitem{Liu:2019zuc} M.~S.~Liu, Q.~F.~L\"{u}, X.~H.~Zhang, and Q.~Zhao,
Phys.\ Rev.\ D \textbf{100}, 016006 (2019). 


\bibitem{Agaev:2023tzi} S.~S.~Agaev, K.~Azizi, B.~Barsbay, and H.~Sundu,
J. Phys. G: Nucl. Part. Phys. \textbf{51}, 115001 (2024).


\bibitem{Agaev:2024pej} S.~S.~Agaev, K.~Azizi, and H.~Sundu,
Phys.\ Lett.\ B \textbf{851}, 138562 (2024). 


\bibitem{Agaev:2024pil} S.~S.~Agaev, K.~Azizi, and H.~Sundu,
Phys.\ Lett.\ B \textbf{856}, 138886 (2024). 


\bibitem{Agaev:2024xdc} S.~S.~Agaev, K.~Azizi, and H.~Sundu,
arXiv:2406.06759 [hep-ph].


\bibitem{Chen:2019vrj} X.~Chen, 
Phys.\ Rev.\ D \textbf{100}, 094009 (2019). 


\bibitem{Bedolla:2019zwg} M.~A.~Bedolla, J.~Ferretti, C.~D.~Roberts, and
E.~Santopinto
Eur.\ Phys.\ J. C \textbf{80}, 1004 (2020). 


\bibitem{Cordillo:2020sgc} M.~C.~Gordillo, F.~De~Soto, and J.~Segovia
Phys.\ Rev.\ D \textbf{102}, 114007 (2020). 


\bibitem{Weng:2020jao} X.~Z.~Weng, X.~L.~Chen, W.~Z.~Deng, and S.~L.~Zhu
Phys.\ Rev.\ D \textbf{103}, 034001 (2021). 


\bibitem{Yang:2021zrc} Z.~H.~Yang, Q.~N.~Wang, W.~Chen, and H.~X.~Chen
Phys.\ Rev.\ D \textbf{104}, 014003 (2021). 


\bibitem{Hoffer:2024alv} J.~Hoffer, G.~Eichmann, C.~S.~Fischer,
Phys.\ Rev.\ D \textbf{109}, 074025 (2024). 


\bibitem{Shifman:1978bx} M.~A.~Shifman, A.~I.~Vainshtein and V.~I.~Zakharov,
Nucl.\ Phys.\ B \textbf{147}, 385 (1979).


\bibitem{Shifman:1978by} M.~A.~Shifman, A.~I.~Vainshtein and V.~I.~Zakharov,
Nucl.\ Phys.\ B \textbf{147}, 448 (1979).


\bibitem{Agaev:2023ara} S.~S.~Agaev, K.~Azizi, B.~Barsbay, and H.~Sundu,
Phys.\ Rev.\ D \textbf{109}, 014006 (2024). 


\bibitem{Jaffe:2004ph} R.~L.~Jaffe, 
Phys.\ Rept.\ \textbf{409}, 1 (2005).


\bibitem{Agaev:2020zad} S.~S.~Agaev, K.~Azizi and H.~Sundu,
Turk.\ J.\ Phys.\ \textbf{44}, 95 (2020). 


\bibitem{PDG:2022} R.~L.~Workman \textit{et al.} [Particle Data Group],
Prog.\ Theor.\ Exp.\ Phys.\ \textbf{2022}, 083C01 (2022).


\bibitem{Godfrey:2004ya} S.~Godfrey,
Phys.\ Rev.\ D \textbf{70}, 054017 (2004). 


\bibitem{Veliev:2010vd} E.~V.~Veliev, K.~Azizi, H.~Sundu, and N.~Aksit,
J.\ Phys.\ G \textbf{39}, 015002 (2012). 


\bibitem{Wang:2024fwc} Z.~G.~Wang,
Chin. Phys. C \textbf{48}, 103104 (2024).


\bibitem{Eichten:2019gig} E.~J.~Eichten, and C.~Quigg,
Phys.\ Rev.\ D \textbf{99}, 054025 (2019). 


\bibitem{Rosner:2015wva} J.~L.~Rosner, S.~Stone, and R.~S.~Van de
Water,(2015) 
arXiv:1509.02220.


\bibitem{Ioffe:2005ym} B.~L.~Ioffe, 
Prog.\ Part.\ Nucl.\ Phys. \textbf{56}, 232 (2006).


\bibitem{Narison:2018dcr} S.~Narison,
Int.\ J.\ Mod.\ Phys.\ A \textbf{33}, 1850045 (2018).
\end{thebibliography}
\end{document}